\documentclass[%
	times%
        ,5p%
        ,twocolumn%
	,sort&compress%
	]{elsarticle}

\usepackage{graphicx}
\usepackage{amssymb}
\usepackage{amsmath} % incompatible with iopart.cls

\usepackage[mathscr]{eucal}

\newcommand{\bbset}[1]{\mathbb{#1}}
\newcommand{\Real}{\bbset{R}}
\newcommand{\Integer}{\bbset{Z}}
\newcommand{\set}[1]{\left\{ #1 \right\}}
\newcommand{\ket}[1]{|{}#1{}\rangle}

\newcommand{\ketbra}[2]{|{}#1{}\rangle\langle{}#2{}|}
\newcommand{\bvec}[1]{\boldsymbol{#1}} % Bold Vector with ams

\newcommand{\MP}{\mathcal{P}} % Projectors
\newcommand{\MM}{\mathcal{M}} % Model/Manifold
\newcommand{\MB}{\mathcal{B}} % base

\newcommand{\MYVERSION}{{\sffamily\textbf{TMU preprint} 2015-03-27}}

\makeatletter
\newcommand{\CustumFooter}[1]{%
  \def\ps@pprintTitle{\let\@oddhead\@empty\let\@evenhead\@empty%
    \def\@oddfoot{\footnotesize{}#1{}\hfill}%
    \let\@evenfoot\@oddfoot}}
\makeatother

\CustumFooter{\MYVERSION}

\begin{document}

\begin{frontmatter} % elsarticle

%%% Comment out below for submission: put revision No.
% \preprint{{\sf 
%     Preprint 2014-09-18
%   }}
%\date{} % Comment out for submission

%%%
%%% Title page
%%%
\title{Bloch vector, disclination and exotic quantum holonomy}

\author[at]{Atushi Tanaka}
\ead[url]{http://researchmap.jp/tanaka-atushi/}

\author[tc]{Taksu Cheon}
\ead[url]{http://researchmap.jp/T\_Zen/} % underscore _ requres \

\address[at]{Department of Physics, Tokyo Metropolitan University,
  Hachioji, Tokyo 192-0397, Japan}
\address[tc]{%
  Laboratory of Physics, Kochi University of Technology, Tosa Yamada, 
  Kochi 782-8502, Japan}

% \author{Atushi Tanaka$^1$ and Taksu Cheon$^2$}
% \address{$^1$ %NOPREPRINT%
%   Department of Physics, Tokyo Metropolitan University,
%   Hachioji, Tokyo 192-0397, Japan}
% \ead{tanaka-atushi@tmu.ac.jp}%NOPREPRINT%
% \address{$^2$ %NOPREPRINT%
%   Laboratory of Physics, Kochi University of Technology, Tosa Yamada, 
%   Kochi 782-8502, Japan} 
% \ead{taksu.cheon@kochi-tech.ac.jp}%NOPREPRINT%

%PREPRINT%\author{Atushi Tanaka}
%\email[]{tanaka-atushi@tmu.ac.jp}
%PREPRINT%\homepage[]{\tt http://researchmap.jp/tanaka-atushi/}
%PREPRINT%\affiliation{Department of Physics, Tokyo Metropolitan University,
%PREPRINT%   Hachioji, Tokyo 192-0397, Japan}

%PREPRINT%\author{Taksu Cheon}
%PREPRINT%%\email[]{taksu.cheon@kochi-tech.ac.jp}
%PREPRINT%%% \tt is a workaround for underscore(_) in \homepage[]{...}
%PREPRINT%\homepage[]{\tt http://researchmap.jp/T_Zen/}
%PREPRINT%\affiliation{Laboratory of Physics, Kochi University of Technology,
%PREPRINT%  Tosa Yamada, Kochi 782-8502, Japan}

%\date{\today}

\begin{abstract}
  A topological formulation of the eigenspace anholonomy, where
  eigenspaces are interchanged by adiabatic cycles, is introduced. The
  anholonomy in two-level systems is identified with a disclination of
  the director (headless vector) of a Bloch vector, which
  characterizes eigenprojectors.
  %{\AT3%
  The covering map structure behind the exotic quantum holonomy
  and the role of the homotopy classification of adiabatic cycles
  are 
  %{\TC4 
  elucidated.
  %} 
  %}%\AT3%
  The extensions of this formulation to
  nonadiabatic cycles and $N$-level systems are outlined.
\end{abstract}

%\pacs{03.65.-w,03.65.Vf,02.40.Pc}
%% PACS2010
%% 03.65.-w 	Quantum mechanics
%% 03.65.Vf     Phases: geometric; dynamic or topological 
%% 02.40.Pc 	General topology 

%PREPRINT% \maketitle

\end{frontmatter} % elsarticle
\section{Introduction}
\label{sec:Introduction}
An adiabatic cycle, where an external parameter of a system is varied
infinitely slowly along a closed path, may induce a nontrivial change,
which is sometimes referred to as an anholonomy. A famous example is
the geometric phase in stationary
states~\cite{Berry-PRSLA-392-45}. This is also called as a phase
holonomy, which is derived from the interpretation in terms of the
differential geometry~\cite{Simon-PRL-51-2167}. Besides the phase of
state vector, it turned out that (quasi-)eigenenergies and eigenspaces
of stationary states of a closed quantum system may exhibit
anholonomies~\cite{Cheon-PLA-248-285}. We call such a change an exotic
quantum holonomy.

A more precise description of the exotic quantum holonomy is the
following. Suppose that a system is initially in a stationary state,
and undergo a unitary time evolution induced by an adiabatic
cycle. The final state of the system is orthogonal to the initial
state. In accordance to the correspondence between eigenstates and
eigenenergies, the trajectory of the eigenenergy connects two
different eigenenergies of the initial system. In other words, the
adiabatic cycle interchanges the eigenspaces and
eigenenergies~\cite{Cheon-PLA-248-285}.

Note that the exotic quantum holonomy is different from Wilczek-Zee's
holonomy~\cite{Wilczek-PRL-52-2111}, since the former occurs even when
there is no spectral degeneracy. Also note that the exotic quantum
holonomy is different from the interchange, or the flip, of
eigenvectors induced by a cycle around a non-Hermitian degeneracy
point~\cite{Uzdin-JPA-44-435302}, which is known as Kato's exceptional
point (EP)~\cite{KatoExceptionalPoint,biorthogonal}, in spite of their
resemblance explained below. This is because the decay due to the
non-Hermiticity makes the stringent limitation to observe the flip in
the adiabatic
limit~\cite{Uzdin-JPA-44-435302,Berry-JPA-44-435303}. Namely, the flip
due to EPs is observable only in the non-unitary time evolution whose
timescale is shorter than the relevant lifetimes in the unstable
system, and, in the parametric
evolution~\cite{Dembowski-PRL-86-787,Lee-PRL-103-134101}.

The earliest example of the exotic quantum holonomy is the
one-dimensional particle under a generalized point
potential~\cite{Cheon-PLA-248-285,Cheon-AP-294-1,Tsutsui-JMP-42-5687}. Since
then, examples are found in quantum
maps~\cite{Tanaka-PRL-98-160407,Miyamoto-PRA-76-042115}, quantum
circuits~\cite{Tanaka-EPL-96-10005}, and quantum
graphs~\cite{Ohya-AP-331-299,Ohya-AP-351-900,Cheon-ActaPolytechnica-53-410}. It
is also shown that the Lieb-Liniger
model~\cite{Lieb-PR-130-15,Ichikawa-PRA-86-015602}, which describes
one-dimensional Bose systems, exhibits the exotic quantum holonomy
along a cycle made of the confinement induced
resonance~\cite{Olshanii-PRL-81-938,Haller-Science-325-1224}, whose
experimental realization should be feasible within the current state
of the art~\cite{Yonezawa-PRA-87-062113}. As an application, an
acceleration of the adiabatic quantum computation was
examined~\cite{Tanaka-PRA-81-022320}.

There are several theoretical works on the exotic quantum
holonomy. Firstly, Fujikawa's gauge theoretical formulation for the
phase holonomy~\cite{Fujikawa-PRD-72-025009,Fujikawa-AP-322-1500} is
extended so as to incorporate the eigenspace anholonomy, which is
understood as a holonomy of basis
vectors~\cite{Cheon-EPL-85-20001,TANAKA-AP-85-1340}. Secondly, Viennot
proposed another gauge theoretical formulation based on the adiabatic
Floquet theory, where the nontrivial Floquet block change is discussed
in terms of gerbes~\cite{Viennot-JPA-42-395302}. Thirdly, the exotic
quantum holonomy is associated to the state flip induced by EPs
through the generalized Fujikawa
formalism~\cite{Kim-PLA-374-1958,Tanaka-JPA-46-315302}. Here the
analytic continuation of a Hermitian (or unitary) adiabatic cycle
provides non-Hermitian cycles, which relate the Riemann surface
structure of eigenenergies~\cite{MehriDehnavi-JMP-49-082105} to the
exotic quantum holonomy.

In spite of these efforts, there still 
%{\AT3%
remain
%remains 
%}%\AT3%
puzzling points on the
nature of the exotic quantum holonomy.  Firstly, the eigenspace
anholonomy and the phase holonomy are mixed in the generalized
Fujikawa formalism, and there is only an ad hoc procedure to
disentangle them~\cite{Tanaka-JPA-45-335305}. Also, it is not
straightforward to extract the geometrical picture from the
generalized Fujikawa formalism.  Although a solution to this problem
is given by Simon's formulation as for the phase
holonomy~\cite{Simon-PRL-51-2167}, there has been no known counterpart
of the Simon's formulation for the exotic quantum holonomy.  Secondly,
Viennot's theory is for periodically driven systems and, is not
applicable to the examples of the exotic quantum holonomy in
autonomous (Hamiltonian)
systems~\cite{Cheon-PLA-374-144,Yonezawa-PRA-87-062113,Cheon-ActaPolytechnica-53-410}.
Thirdly, the association of the exotic quantum holonomy to the
non-Hermitian degeneracy requires the analyticity of the adiabatic
parameter. However, it is easy to construct the examples that lack the
analyticity~\cite{Cheon-ActaPolytechnica-53-410}. Hence the EP
interpretation of the exotic quantum holonomy is applicable only to a
limited class of examples.

% AT
In this manuscript, we introduce a topological formulation of the exotic quantum holonomy, in particular the eigenspace anholonomy, to clarify these problems.
This enables us to identify the quantity 
%{\AT3%
that
%which 
%}%\AT3%
predicts
whether a given cycle exhibits the eigenspace anholonomy. 
%
%AT%In this manuscript we develop a topological formulation for 
%AT%the eigenspace anholonomy,  
%AT%%TC in order to identify 
%AT%that enables to identify the quantity which predicts
%AT%whether a given cycle exhibits the eigenspace anholonomy. 
%AT%It turns out that the present formulation naturally lead to
%AT%a nonadiabatic extension of the exotic quantum holonomy.
%
%The aim of this manuscript is to introduce a topological formulation for 
%the eigenspace anholonomy. 
%
%TC First, we  employ
The first key concept in our approach is
{\it the ordered set of eigenprojectors}.
%TC (Eq.~\eqref{eq:p_def}).
We regard the eigenspace anholonomy as a permutation, which is induced
by an adiabatic cycle, among the elements of the ordered set of 
eigenprojectors.
This may be considered as a counterpart to Simon's vector
bundle formulation of the phase holonomy~\cite{Simon-PRL-51-2167}.
%An adiabatic cycle permutates the elements of the ordered set of 
%eigenprojectors. 
In a two-level system, we show that a permutation occurs only when 
the adiabatic cycle encloses a singular point odd times.
The second key concept in our topological formulation is {\it the cycles in
a quantum dynamical variable} 
%TC (Eq.~\eqref{eq:b_def}) 
that 
%{\AT3%
takes
%take 
%}%\AT3%
the place of conventional 
cycles in adiabatic parameters.
It is shown that the topological nature of the eigenspace anholonomy
has a direct link with the homotopy classification of 
cycles~\cite{Mermin-RMP-51-591}.
For example, the singular point mentioned above may be called 
a disclination, or line defect~\cite{Director}. 
%TC (``line'' defect)
%At the same time, we explain that the homotopy of the cycles governs 
%the permutation among eigenprojectors.
%
Furthermore, 
in common with
%TC Aharonov and Anandan's 
% AT
Aharonov-Anandan formulation~\cite{Aharonov-PRL-58-1593},
%AT Aharonov-Anandan formulation,
%TC ~\cite{Aharonov-PRL-58-1593} for the phase holonomy,
the new definition of the cycles 
allows us to extend the eigenspace anholonomy to nonadiabatic cycles.

\begin{figure}%[bt]
  \centering
  \includegraphics[width=3.5cm]{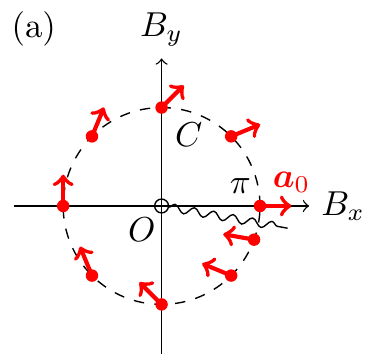}  
  \includegraphics[width=3.5cm]{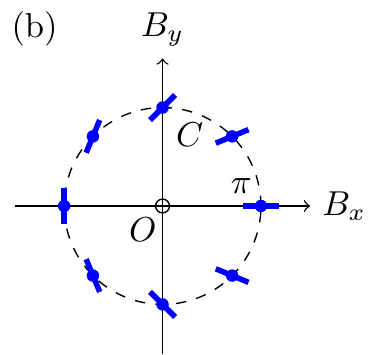}  
  \caption{
    A disclination of eigenobjects of quantum kicked spin-$\frac{1}{2}$
    (Eq.~\eqref{eq:U_def}) in the $(B_x,B_y)$-plane.
    (a)~The Bloch vector $\bvec{a}$ (Eq.~\eqref{eq:def_a}) 
    at a circle $C$ ($|\bvec{B}|=\pi$).
    %AT2
    %The filled circle at $(2\pi,0)$ is a diabolic point.
    The Bloch vector is not well-defined at the origin, which 
    reflects the multiple-valuedness of $\bvec{a}$.
    A ``branch cut'' is depicted by a wavy line.
    Here $\bvec{a}_0=\bvec{e}_x$ is the normalized Bloch vector
    at the initial point $(\pi, 0)$ of $C$.
    The adiabatic time evolution along $C$ induces a flip of $\bvec{a}$.
    (b)
    The director (headless vector) $\bvec{n}$ of $\bvec{a}$
    at the circle $C$. Since $\bvec{n}$ is single-valued in
    the $(B_x,B_y)$-plane, no branch cut needs to be drawn. Still,
    the line defect remains at the origin.
    }
  \label{fig:an_schematic}
\end{figure}

\section{Quantum kicked spin-$\frac{1}{2}$ with
  two adiabatic parameters}
Throughout 
%TC the explanation of the present 
the presentation of our formulation, we assume 
that the systems are described either by Hermitian 
%AT4
Hamiltonians
%Hamiltonian 
or by
unitary Floquet 
%AT4
operators.
%operator.
%TC
We also assume that there is no spectral degeneracy 
in the adiabatic cycles. 
For concreteness, we employ a periodically driven spin-$\frac{1}{2}$
%TC to explain 
to illustrate our formulation, which is immediately applicable to an 
arbitrary two-level system. 
An extension to an arbitrary $N$-level 
%{\AT3%
system
%systems 
%}%\AT3%
is also to be shown.
%TC below.

Let us suppose that
%TC We suppose that 
%We introduce the point above through an analysis of 
%the quantum kicked spin-$\frac{1}{2}$
%the system has
%which 
the system is described by a time-periodic 
%AT2: add an endnote
Hamiltonian\footnote{%
  The exotic quantum holonomy in an autonomous
  Hamiltonian system requires either a level
  crossing or the divergence of
  eigenenergy~\cite{Cheon-PLA-374-144}. To avoid
  complications from such singularities, we here 
  examine periodically driven systems.}:
$
  \hat{H}(t)
  \equiv 
  \frac{1}{2} 
  \bvec{B}\cdot\hat{\bvec{\sigma}} 
  + \frac{1}{2}
  %{\AT3%
    \lambda
    %\phi 
  %}%\AT3%
  ({1 -\hat{\sigma}_z})
  \sum_{m=-\infty}^{\infty}\delta(t-m)
  ,
$
%TC where $\hat{\sigma}_j$ ($j=x,y,z$) is the Pauli matrix,
%and we set 
%$\hat{\bvec{\sigma}} 
%\equiv
%\bvec{e}_x\hat{\sigma}_x+\bvec{e}_y\hat{\sigma}_y+\bvec{e}_z\hat{\sigma}_z$.
where 
%{\AT3%
  $\bvec{B}$ is a static magnetic field
  and
%}%\AT3
$\hat{\bvec{\sigma}} 
\equiv
\bvec{e}_x\hat{\sigma}_x+\bvec{e}_y\hat{\sigma}_y+\bvec{e}_z\hat{\sigma}_z$ is a
unimodular linear combination of Pauli matrices
$\hat{\sigma}_j$ ($j=x,y,z$).
%and
%$\bvec{e}_{\rho} 
%\equiv \bvec{e}_x\cos\phi+\bvec{e}_y\sin\phi$.
%
The Hamiltonian $\hat{H}(t)$ contains a periodically
pulsed rank-$1$ perturbation~\cite{Combesqure-JSP-59-679}
with strength 
%{\AT3%
$\lambda$.
%$\phi$.
%}%\AT3%

%{\AT3%
In the following, $\bvec{B}$ 
%{\AT4
is
%}%\AT4
%{\TC4 
assumed to be restricted within $xy$-plane,
i.e., $\bvec{B}=B_x\bvec{e}_x+B_y\bvec{e}_y$.
%}
To parameterize $\bvec{B}$ with the cylindrical coordinates $B$ and $\phi$,
we introduce
$\bvec{e}_{\rho} \equiv \bvec{e}_x\cos\phi+\bvec{e}_y\sin\phi$
and 
$\bvec{e}_{\phi} \equiv \bvec{e}_y\cos\phi-\bvec{e}_x\sin\phi$.
Hence we have $\bvec{B} = B\bvec{e}_{\rho}$.
We also impose $\lambda=\phi$ in the following.
%Hence the adiabatic parameters of this models are $(B,\phi)$, or
%equivalently, $(B_x,B_y)$.
%}%\AT3%
We introduce a Floquet operator, which describes a unit time evolution
generated by $\hat{H}(t)$:
\begin{equation}
  \label{eq:U_def}
  \hat{U}
  \equiv 
  \exp\left(-i\phi\frac{1 -\hat{\sigma}_z}{2}\right)
  \exp\left(-\frac{i}{2}
    %{\AT3%
    B\bvec{e}_{\rho}%
    %\bvec{B}
    %}%\AT3%
    \cdot{\hat{\bvec{\sigma}}}\right)
  .
\end{equation}
%TC
It is straightforward to show that $\hat{U}$ is periodic in $\phi$ with
the period $2\pi$~\cite{Tanaka-PRL-98-160407,TANAKA-AP-85-1340}.
%TC (ADD ENDNOTE).
Accordingly we identify the 
%{\AT3%
adiabatic
%}%\AT3%
parameter space of the model with
a two-dimensional plane $(B_x, B_y)\equiv(B\cos\phi, B\sin\phi)$.

We now diagonalize $\hat{U}$. First, 
$\hat{U}$ is expanded as 
%we expand $\hat{U}$ by $\hat{\bvec{\sigma}}$:
%\begin{equation}
$
  \hat{U}
  %&
  =
  e^{-i\phi/2}
  \left[
    \cos({\Delta}/{2})
    -i\hat{\bvec{\sigma}}\cdot\tilde{\bvec{a}}
  \right]
  ,
$
%\end{equation}
where
\begin{eqnarray}
  \Delta
  &
  \equiv& 2\arccos\left(\cos\frac{\phi}{2}\cos\frac{B}{2}\right)
  ,\\
  \tilde{\bvec{a}}
  &
  \equiv&
  \left(\bvec{e}_{\rho}\cos\frac{\phi}{2}
    - \bvec{e}_{\phi}\sin\frac{\phi}{2}\right)
  \sin\frac{B}{2}
  \nonumber\\&&\quad{}
  -\bvec{e}_z\sin\frac{\phi}{2}\cos\frac{B}{2}
  .
  %,
\end{eqnarray}
Because $\tilde{\bvec{a}}\cdot\tilde{\bvec{a}} = \sin^2({\Delta}/{2})$ holds,
the eigenvalues of $\hat{U}$ 
%TC degenerates where
become degenerate when
$\sin({\Delta}/{2}) = 0$ holds. 
%TC We exclude
Excluding the degeneracy points $\bvec{B} = \bvec{0}, 2\pi\bvec{e}_x, 4\pi\bvec{e}_x, \dots$,
%TC to examine the eigenspace anholonomy in the following.
%Hence it is legitimate to
we can normalize $\tilde{\bvec{a}}$:
\begin{equation}
  \label{eq:def_a}
  \bvec{a}
  \equiv
  \tilde{\bvec{a}} / \sin({\Delta}/{2})
  %\left({\sin\frac{\Delta}{2}}\right)^{-1}\tilde{\bvec{a}}
  .
\end{equation}
%TC Accordingly 
We obtain the spectral decomposition of $\hat{U}$ in the form
% in terms of $\bvec a$ as
\begin{equation}
  \label{eq:spectral_decomposition}
  \hat{U}
  = z_+
  \hat{P}(\bvec{a}) 
  + z_-
  \hat{P}(-\bvec{a}) 
  ,
\end{equation}
where
$z_{\pm}\equiv e^{-i(\phi \pm\frac{\Delta}{2})}$ are eigenvalues, and
\begin{equation}
  \hat{P}(\bvec{a}) 
  \equiv \frac{1 +\bvec{a}\cdot\hat{\bvec{\sigma}}}{2}
  ,
\end{equation}
is a projection operator parameterized by a unit vector $\bvec{a}$,
which is called a (normalized) Bloch vector.
Eq.~\eqref{eq:spectral_decomposition}
implies that $\hat{P}(\pm\bvec{a})$ are the eigenprojectors of $\hat{U}$.
In other words, 
%TC
for a given pair of $B_x$ and $B_y$, except at the degeneracy points,
there are two normalized Bloch vectors $\pm\bvec{a}$, which 
correspond to two eigenprojectors. 
We note that the spectral decomposition 
(Eq.~\eqref{eq:spectral_decomposition}) is applicable to
an arbitrary unitary Floquet operator or Hermitian Hamiltonian
as long as the corresponding two level system has no spectral degeneracy.
Hence the following argument is applicable to two-level systems
in general.

%{\AT3%
%{\TC4 
Prior to the consideration of the exotic quantum holonomy in the kicked spin~\eqref{eq:U_def}, a remark is due to
%}
%induced by adiabatic cycles, 
%we remark on 
the adiabatic time evolution in periodically driven systems,
%Here the
%{\TC4 
for which
%}
parametric evolution of an eigenvector of the Floquet operator
$\hat{U}$ describes the adiabatic time evolution,
%{\TC4 
in place of 
%{\AT4%
a
%the
%}%\AT4%
Hamiltonian,
%{\ATDEL4{}$\hat{H}(t)$}%
%,
%}%
up to a phase factor. 
The dynamical phase is
determined by a quasienergy~\cite{Zeldovich-JETP-24-1006} 
%{\TC4 
in place of
%}
eigenenergy. Proofs of the adiabatic theorem for Floquet systems are
found in
Refs.~\cite{Young-JMP-11-3298,Dranov-JMP-39-1340,Tanaka-JPSJ-80-125002}.
The corresponding adiabatic condition is governed by the gaps of
quasienergies 
%{\TC4 
in place of
%}
eigenenergies~\cite{Breuer-ZPD-11-1}.
%}%\AT3%

We examine the adiabatic time evolution of the eigenprojector
$\hat{P}(\bvec{a})$ along a cycle $C$
in the $(B_x, B_y)$-plane.
%TC
It is sufficient to examine the evolution of the normalized Bloch vector
$\bvec{a}$ instead of $\hat{P}(\bvec{a})$ due to their equivalence.
We depict the parametric evolution of $\bvec{a}$ in 
Fig.~\ref{fig:an_schematic} (a).
Let $\bvec{a}=\bvec{a}_0$ at the initial point $\bvec{B}_0$ 
%TC 
on $C$.
After a completion of the counterclockwise adiabatic rotation of $\bvec{B}$
%TC in the counterclockwise direction 
along $C$, $\bvec{a}$ arrives
at $-\bvec{a}_0$ (see, Fig.~\ref{fig:an_schematic}(a)),
which implies that the final eigenprojector $\hat{P}(-\bvec{a}_0)$
is orthogonal to the initial one $\hat{P}(\bvec{a}_0)$.
Hence $C$ induces the interchange of eigenprojectors $\hat{P}(\pm\bvec{a}_0)$
%TC to realize 
resulting in the realization the eigenspace anholonomy.
%TC
This fact is stable against the deformation of
the adiabatic cycle, as long as $C$ encloses the origin only once.
%TC On the other hand, 
Let us next examine the case that $C$ does not
enclose the origin. 
%TC
The simplest case is the one where $C$ start
from $\bvec{B}_0$ and keeps to stay $\bvec{B}_0$, i.e., $C$ is a trivial
cycle. The direction of the Bloch vector at the final point of the cycle 
agrees with the one at the initial point. Namely, the eigenprojector 
returns to the original one after the completion 
of the adiabatic cycle. 
This remains correct as long as $C$ does not enclose the origin $O$
in the $(B_x, B_y)$-plane.
%TC
Also, the initial and final Bloch vectors are the same when $C$ encloses
the origin even times.

\section{%
  Eigenspace anholonomy
  as an anholonomy of
  an ordered set of mutually orthogonal projection operators
}
%TC
Here, we propose a novel interpretation of the normalized Bloch vector
$\bvec{a}$ 
%TC to investigate the eigenspace anholonomy in order to extend 
that allows the extension of our analysis to systems with an arbitrary 
number of levels.
%
%TC In the following, one of 
The central object is 
%TC {\em 
an ordered set of mutually orthogonal projection operators
%}
\begin{equation}
  \label{eq:p_def}
  p\equiv
  \bigl(|\psi_0\rangle\langle\psi_0|, |\psi_1\rangle\langle\psi_1|\bigr)
 , 
\end{equation}
which can be specified by a normalized Bloch vector
\begin{equation}
  p(\bvec{a})~= (\hat{P}(\bvec{a}),\hat{P}(-\bvec{a}))
  .
\end{equation}
%TC
A given pair of $B_x$ and $B_y$, except at the degeneracy points,
specifies two normalized Bloch vectors $\pm\bvec{a}$.
One of them, say $\bvec{a}$,
precisely determines $p$. Another normalized Bloch vector $-\bvec{a}$
correspond to another ordered set of projection operators
$p(-\bvec{a})=(\hat{P}(-\bvec{a}),\hat{P}(\bvec{a}))$, which is obtained
by a permutation of the elements of $p(\bvec{a})$.
%
%TC
As for two-level systems, we can identify $p$ 
with a normalized Bloch vector $\bvec{a}$, 
%in order to fully utilize 
which helps our geometric intuition. 
%TC In other words,
The $p$-space of two-level systems is equivalent to the sphere $S^2$.

In terms of the ordered set of projectors $p$, 
the eigenspace anholonomy is the permutation of the elements
of $p$ induced by an adiabatic cycle.
For example, let us start an adiabatic cycle $C$ that enclose the origin
of $(B_x, B_y)$-plane in  Fig.~\ref{fig:an_schematic}(a).
After the completion of the cycle $C$, the elements of $p$ are interchanged.
In other words, $C$ corresponds to a permutation of the elements of $p$.
%

%%%%%%%%%%%%%%%%%%%%%%%%%%%%%%%%%%%%%%%%%%%%%%%%%%%%%%%%%%%%%%%%%%%%%%%%%%%%
%{\AT3%
%{\tt Covering map from Bloch sphere to adiabatic parameter space}

%{\TC4 
In this study, we make use of topological concepts of
%}
% {\ATDEL4%
% the 
% }%ATDEL4{%
covering maps and
% {\ATDEL4%
% the 
% }%ATDEL4{%
% Following Mermin's RMP
homotopy
%{\AT4%
groups~\cite{Lee-ITT-2011,Nakahara-GTP-1990}.
%}%\AT4%
%{\TC4 
These concepts
%} 
have been
utilized, for example, in the studies of topological
defects~\cite{Mermin-RMP-51-591} and exceptional
points~\cite{MehriDehnavi-JMP-49-082105}.  A concise summary of the
covering map is available in Ref.~\cite{MehriDehnavi-JMP-49-082105}.
It will be shown that the resultant interpretation of the eigenspace
anholonomy resembles Simon's interpretation of the phase holonomy in
terms of vector bundles~\cite{Simon-PRL-51-2167}.  Although the
following 
%{\TC4 
description
%} 
is for the two-level kicked spin, the
generalization to systems with an arbitrary number of levels is
straightforward, as 
%{\TC4 
to be explained later.
%}.

%First of all, we explain 
%{\TC4 
Let us be more precise on the relationship between the adiabatic
parameter space and $p$-space.
%}. 
Let $\MM$ denote the
adiabatic parameter space where the spectral degeneracy points are
excluded.  For a given point
%TC4, say 
$x$ in $\MM$, let $\MP_{x}$ denote
the set of two possible values of $p$. Since $\MP_{x}$ may be regarded
as a fiber, we denote the corresponding fiber bundle as $\MP$. The
projection from $\MP$ to $\MM$ is known as a covering map in
topological analysis of
manifolds~\cite{Lee-ITT-2011,MehriDehnavi-JMP-49-082105}.

%Although the appearance of the fiber bundle in the present formulation suggests 
%{\TC4 
Despite the apparent similarity, there are several crucial differences between the
fiber bundle appearing here and the fiber bundle interpretation of
the phase holonomy.
%} 
Among them, we
%{\TC4 
point out the discreteness
%} 
of the structure group of the fibers for
the eigenspace anholonomy. 
%{\TC4 
This comes from the fact that the fiber bundle originates
from the covering map~\cite{Lee-ITT-2011}, so that $p$ is ``quantized''
at a given point in $\MM$.
%}  
%On the other hand, 
%{\TC4 
Contrarily,
%}, 
the structure group for
the phase holonomy is mostly 
%{\TC4 
continuous~\cite{Bohm-GPQS-2003}.
%}

%TC4As for 
For the kicked spin~\eqref{eq:U_def}, $\MM$ is $(B_x, B_y)$-plane
excluding the degeneracy points. For a given point, say $\bvec{B}$, in
$\MM$, there are two Bloch vectors, say $\pm\bvec{a}$, so that a fiber
$\MP_{\bvec{B}}$ consists of the two points $(\bvec{B},
p(\pm\bvec{a}))$. See, figure~\ref{fig:cover2}~(a). The corresponding
structure group is the symmetric group of two elements (i.e., the
group that contains all permutations for two items), and coincides
with $\Integer_2$.
%
%{\AT4%
In order to simplify the following argument, 
we restrict $\MM$ to be an annulus whose center is the origin
in $(B_x, B_y)$-plane
(e.g., $\frac{\pi}{2} < B < \frac{3\pi}{2}$).
(See, figures~\ref{fig:an_schematic}(a) and ~\ref{fig:cover2}~(a)).
%}%\AT4%

%Secondly, 
%{\TC4 
We now
%} 
associate the time evolution of $p$ along an adiabatic
cycle $C$ 
%{\TC4 
to
%} 
the concept of lifting.  For a given initial condition
of $p$, the trajectory of $p$ induced by $C$ essentially determines
the lifting
$\tilde{C}$ of $C$ to $\MP$~\cite{Bohm-GPQS-2003}.
%{\TC4 
The situation is visualised in Fig.~\ref{fig:cover2} (a).
%}.
%
As for the kicked spin, for a given initial point of $C$, say,
$\bvec{B}_0$, there are two possible normalized Bloch vectors
$\pm\bvec{a}_0$, which provides two different lifts of $C$.  Suppose
$\bvec{B}_0$ is slightly varied.  Correspondingly, the Bloch vectors
$\pm \bvec{a}_0$ are smoothly and slightly deformed.  The repetition
of this procedure determines the two lifts $\tilde{C}_{\pm}$ of $C$.

The lift $\tilde{C}$ tells us whether the
eigenprojectors in $p$ are interchanged by the adiabatic cycle
$C$. When $\tilde{C}$ is a closed path, each eigenprojector in $p$
%{\TC4 
also
%} 
draws a closed path. This implies the absence of the eigenspace
anholonomy. On the other hand, when $\tilde{C}$ is open, the initial
and the final points of an eigenprojector are different, which is the
case 
%{\TC4 
in which
%} 
the eigenspace anholonomy occurs.  Furthermore, the
destination of $\tilde{C}$ precisely describes the permutation among
the eigenprojectors.

Let $\phi_C(p)$ denote the final point of the lifted path $\tilde{C}$
of $C$, where $p$ is the initial point of $\tilde{C}$ path. 
%Thus 
Our question of the eigenspace anholonomy is 
%{\TC4 
now
%} 
casted into the problem 
%{\TC4 
of
%}
determining $\phi_C$.
%TC4 , which is an arbitrary permutation.  
%{\TC4  The permutation} 
The mapping
$\phi_C$ is called a monodromy action in the analysis of covering
map~\cite{Lee-ITT-2011}.  For a given initial point in $\MP$, let
$\Phi$ denote the set of $\phi_C$ generated by all possible adiabatic
cycles.  $\Phi$ is called the automorphism group of the covering~\cite{Lee-ITT-2011}, and is the
%TC4  in the studies of covering groups
counterpart of the holonomy group for the phase holonomy.

Not every detail
%TC4s
of $C$ is required to determine $\phi_C$.
%{\TC4 
This is to be expected
%} 
from the fact that $\phi_C$ corresponds to a permutation, and thus is ``quantized''.  
In the analysis of covering map, the
homotopy classification of $C$ plays the central role~\cite{Lee-ITT-2011,MehriDehnavi-JMP-49-082105}.  We say that a
path $C$ is homotopic to another path $C'$, when $C$ can be smoothly
deformed to $C'$ with the initial and final points kept unchanged.
Let $[C]$ denote the class of paths that are homotopic to
$C$~\cite{Lee-ITT-2011}.  Since $\phi_C$ and $\phi_{C'}$ are the same as
long as $C$ is homotopic to $C'$ 
because of
the homotopy lifting property, the quantity $\phi_{C}$ may be denoted as
$\phi_{[C]}$~\cite{Lee-ITT-2011}.
Hence, it suffices to examine $\phi_{[C]}$ 
with $[C]$ belonging to the first fundamental group of $\MM$, {\it i.e.}, $[C]\in\pi_1(\MM)$.

It is important, at the same time, to observe that 
$\phi_{[C]}$ describes the identical permutation
if and only if the lifted cycle $\tilde{C}$ of $C$ is homotopic to 
a closed path in $\MP$, i.e., 
$[\tilde{C}]\in\pi_1(\MP)$.

In order to express this condition in terms of $\MM$,
we consider the projection of all the elements $\pi_1(\MP)$ to $\MM$.
We denote the resultant set as $H$, which is a subset of $\pi_1(\MP)$
and is called an isotropy group.
The quotient space $\pi_1(\MM)/H$ 
precisely classifies $\phi_{[C]}$,
i.e., $\Phi\simeq\pi_1(\MM)/H$, according to the covering
automorphism group structure theorem (Ref~\cite{Lee-ITT-2011}, Theorem
12.7).

For example, the kicked spin~\eqref{eq:U_def} 
has 
$\pi_1(\MM) = \{[e], [\alpha], [\alpha^2], \ldots\}$,
where $e$ is homotopic to a point and $\alpha$ encloses 
the origin $O$ once in $\MM$ (see, figure~\ref{fig:cover2}).
On the other hand, 
$\pi_1(\MP) = \{[e'], [\beta], [\beta^2], \ldots\}$
where $e'$ is homotopic to a point, and $\beta$ encloses 
the ``hole'' in $\MP$.
Because the projection of $\beta$ to $\MM$ is homotopic to $\alpha^2$,
we find $H=\{[e], [\alpha^2], [\alpha^4], \ldots\}$.
Hence, $\Phi=\pi_1(M)/H$ consists of two classes. One 
%{\AT3%
corresponds
%correspond 
%}%\AT3
to
the cycles that 
encloses 
%{\AT3%
enclose
%encloses 
%}%\AT3%
$O$ even times, and makes $\phi_{[C]}$
the identity. The other is composed by
the cycles that 
%{\AT3%
enclose
%encloses 
%}%\AT3%
$O$ odd times to make $\phi_{[C]}$
the cyclic permutation of the two items in $p$.
Hence we conclude that $\Phi$, the automorphism group of the covering,
is equivalent with $\Integer_2$, 
the cyclic group whose order is $2$.
%%
%% AT3: Put some summary here?
%%
%TC 
%In hindsight, it is to be expected, from the topological nature of the problem, 
%that the condition for the eigenspace anholonomy involves the homotopy of adiabatic cycles.
%
%}%\AT3%
%%%%%%%%%%%%%%%%%%%%%%%%%%%%%%%%%%%%%%%%%%%%%%%%%%%%%%%%%%%%%%%%%%%%%%%%%

\section{%
  Definition of cycles by
  quantum dynamical variables
  % a set of mutually orthogonal projection operators
  instead of c-number
  %adiabatic
  parameters
}%
So far, cycles are parameterized by the adiabatic parameters $(B_x, B_y)$.
This has been a common definition in the previous studies of the exotic 
quantum holonomy~\cite{Cheon-PLA-248-285,Cheon-EPL-85-20001}.
Instead, we propose a way to define the cycles  
only in terms of quantum dynamical variables. 
The aim here is twofold. One is to complete a geometrical
view of the eigenspace anholonomy. Another is to extend the exotic quantum
holonomy into nonadiabatic 
%{\AT3%
cycles, which will be examined in the next section.
%cycles.
%}%\AT3%
%
To achieve this, we introduce 
{\em a set of mutually orthogonal eigenprojectors}
\begin{equation}
  \label{eq:b_def}
  b
  \equiv
  \{|\psi_0\rangle\langle\psi_0|, |\psi_1\rangle\langle\psi_1|\} 
 ,
\end{equation}
where the order of the projector are disregarded.

As for the two level systems, we obtain 
a geometric interpretation of $b$
with the help of a normalized Bloch vector $\bvec{a}$
\begin{equation}
 b(\bvec{a})
 =
 \{\hat{P}(\bvec{a}),\hat{P}(-\bvec{a})\}
 ,
\end{equation}
which 
%AT4
agrees
%agree 
with $b(-\bvec{a})$, since the order of the elements in 
$b$ is ignored.
In other words, 
we identify $\bvec{a}$ and $-\bvec{a}$ in the specification of $b$.
%TC
In geometry, the identification of antipodal points on the sphere $S^2$ leads to
the real 
projective plane $\Real{}P^2$~\cite{Mermin-RMP-51-591,Nakahara-GTP-1990}.
Hence we identify $b$ with a point, which we denote as $\bvec{n}$,
in the projective plane.
In Fig.~\ref{fig:an_schematic} (b), 
$\bvec{n}$ is depicted in the $(B_x, B_y)$-plane.
We note that $\bvec{n}$ is single-valued here. Still, we have
a singularity at the origin $O$, where the value of $\bvec{n}$ 
cannot be determined. 
This resembles a disclination of nematic liquid crystals~\cite{Director}.
In the studies of nematic liquid crystals,
$\bvec{n}$ is called as a {\it director}, 
%TC ("headless vector")~\cite{Director}.
or a headless vector~\cite{Director}.
%{\AT3%
The disclination in nematics is a discontinuity of $\bvec{n}$-field
in three-dimensional space and the discontinuity may extends along a 
%{\AT4
line~\cite{Mermin-RMP-51-591,Nakahara-GTP-1990}.
%}%\AT4
In this sense, the disclination is distinct from 
the effective monopole induced by the phase holonomy~\cite{Berry-PRSLA-392-45}.
% {\ATDEL4%
% Note that the discontinuity of this model~\eqref{eq:U_def}
% do not occupy along a line and is confined within the degeneracy points
% in the $(B_x, B_y)$-plane.
% }%\ATDEL4%
%}%AT3%
We use $\bvec{n}$, or equivalently $b$, to define an adiabatic
cycle $C$. 
%TC As 
For the quantum kicked spin~(Eq.~\eqref{eq:U_def}),
we regard that the path $C$ resides 
%is drawn 
in the $\bvec{n}$-space rather than in 
the $(B_x, B_y)$-plane, 

%%%%%%%%%%%%%%%%%%%%%%%%%%%%%%%%%%%%%%%%%%%%%%%%%%%%%%%%%%%%%%%%%%%%%%%%%%%%%
%{\AT3%
It is possible to examine all two-level systems that have no spectral 
degeneracy with $\Real{}P^2$. This is because
the director space $\Real{}P^2$ can
parameterize an arbitrary adiabatic cycle of two level systems 
%{\AT4%
including 
%}%\AT4%
the kicked spin-$\frac{1}{2}$,
as long as the adiabatic cycle do not encounter any spectral degeneracy.
We have a complete classification of the adiabatic cycles 
of non-degenerate two-level systems in following manner.

\begin{figure}[h]
  \centering
  % 3.5cm \simeq 0.23\textwidth 
  \includegraphics[width=4.3cm]{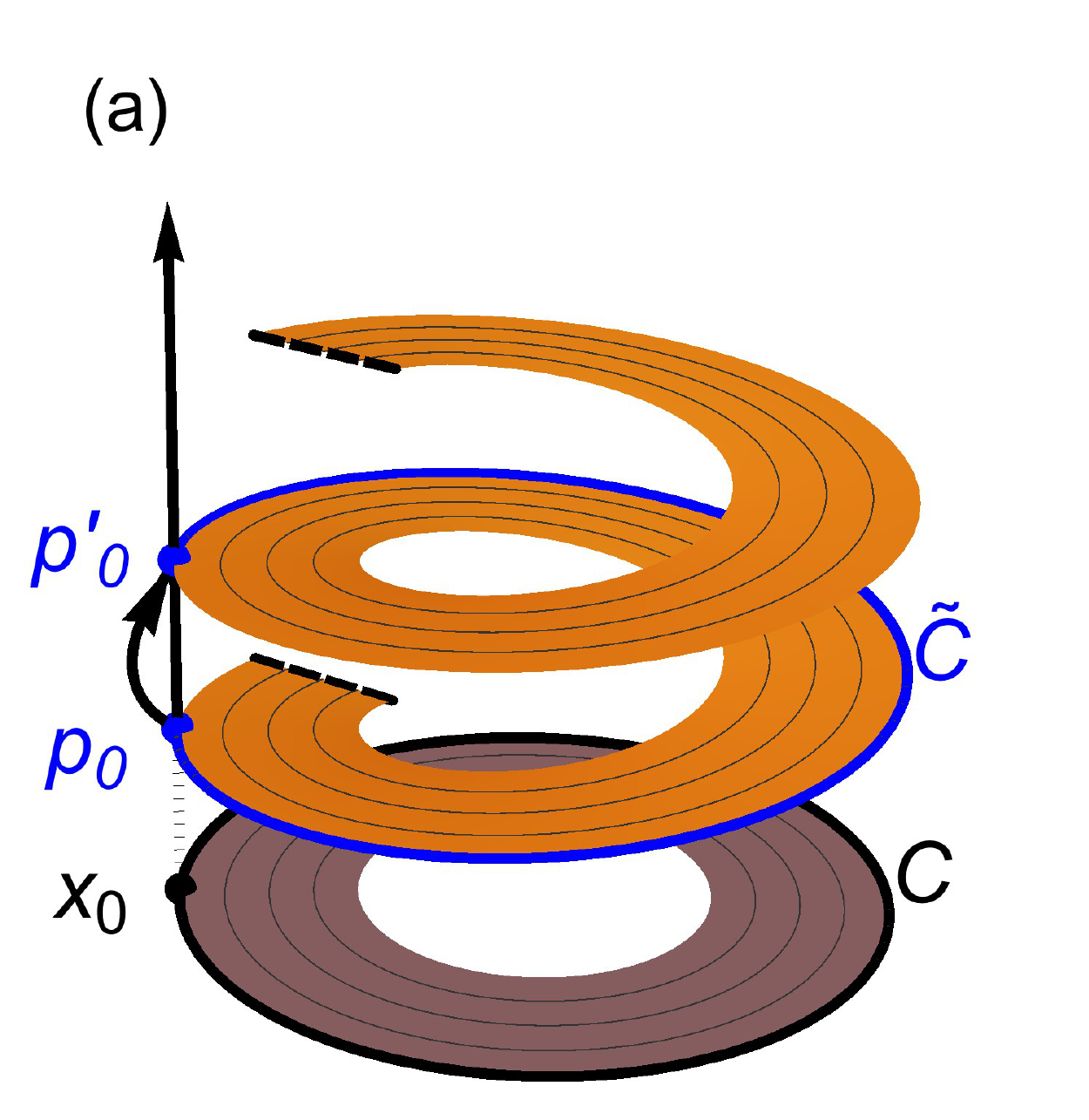}
  \includegraphics[width=3.5cm]{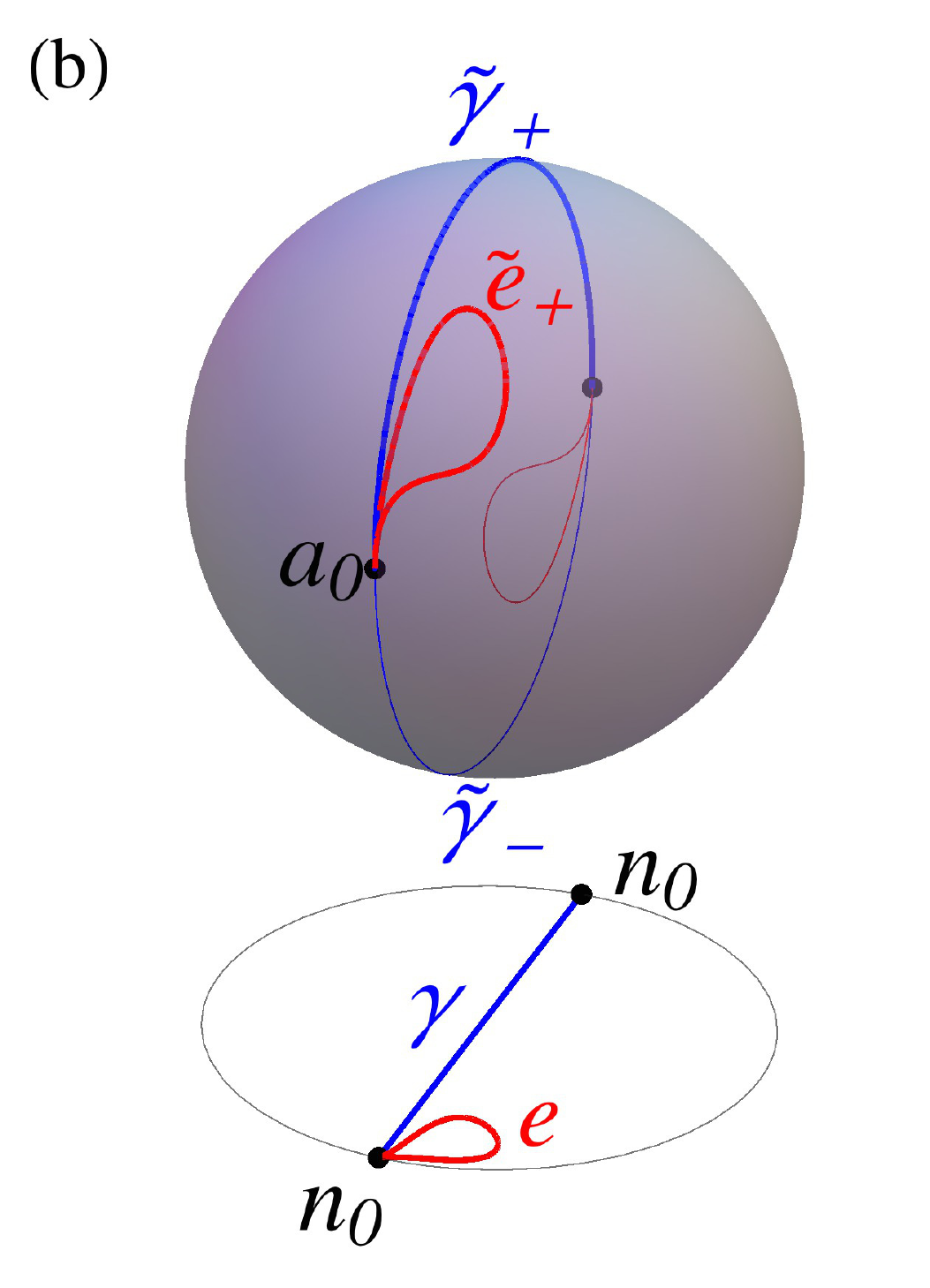}  
  \caption{
    %{\AT3%
      The covering map structures of two-level systems (schematic).
    %}%\AT3%
    (a) 
    %{\AT3%
      The adiabatic space $\MM$ (bottom annulus)
      of the kicked spin-$\frac{1}{2}$~\eqref{eq:U_def}
      is chosen so as to avoid degeneracy points
      (e.g., $\frac{\pi}{2} < B < \frac{3\pi}{2}$ in Figure~\ref{fig:an_schematic}).
      Since two edges depicted by dashed lines of $\MP$ (the winding strip
      in the above)
      should be identified, $\MP$ is a ``double-winding'' strip above $\MM$.
      The projection from $\MP$ to $\MM$ is a covering map.
      The adiabatic cycle
      $C$ starts from $x_0$ and winds $\MM$ in the counterclockwise direction.
      The corresponding lifting $\tilde{C}$ delivers $p_0$ to $p'_0$.
      Accordingly the monodromy action $\phi_C$ describe the motion
      in the fiber direction.
    %}%\AT3%
       (b) 
    %{\AT3%
      Adiabatic cycles of the two-level systems without spectral
      degeneracies can be parameterized by the projective plane $\MM$,
      which is equivalent with $\Real{}P^2$,
      shown in the bottom part. 
      Two cycles (closed paths) $e$ and $\gamma$ are shown there.
    %}%\AT3%
        Two filled circles corresponding to $\bvec{n}_0$, which is the
    initial point of these cycles, 
    are identical in $\Real{}P^2$.
    %are an identical point in $\Real{}P^2$.
    The cycle $e$ is homotopic to a zero-length
    cycle, and is not homotopic to $\gamma$.
    In the top, 
    the lifts of $e$ and $\gamma$ to the $p$-space,
    which is equivalent to $S^2$, are shown.
    %which is equivalent with $S^2$ as for the two-level systems, are shown.
    Because $S^2$ doubly covers $\Real{}P^2$,
    there are two normalized Bloch vectors $\pm\bvec{a}_0$ for 
    a given director $\bvec{n}_0$.
    Also, each cycle has two lifts (thick and dashed curves).
    The lifts $\tilde{e}_{\pm}$ of $e$ are closed, 
    signifying
    % which correspond to 
    the absence of eigenspace anholonomy.
    On the other hand, the lifts $\tilde{\gamma}_{\pm}$ are open.
    Along the adiabatic cycle $\gamma$, 
    the initial point $\bvec{a}_0$ of $\tilde{\gamma}_{+}$
    is transposed to $-\bvec{a}_0$, which is the initial point of 
    $\tilde{\gamma}_{-}$, and vice versa.
    }
  \label{fig:cover2}
\end{figure}

Let $\MB$ denote the adiabatic parameter space $\Real{}P^2$.
Now $\MP$ is equivalent to the whole sphere $S^2$.
There is a covering map from $\MP$ to $\MB$, in the sense that
the inverse of the covering map of $\bvec{n}$ is 
the set of two Bloch vectors $\pm\bvec{n}$.
The fundamental group of $\MP$ is $\pi_1(\MP) = \set{[e']}$, 
where $e'$ is a cycle homotopic to a point in $\MP$.
%Hence 
%This is because 
The simplicity of $\pi_1(\MP)$ compared to the previous case is the reflection of
the fact that $\MP$ is ``larger'' than 
%{\AT4%
the
%}%\AT4%
previous one, as we have
``extended'' the base space from $\MM$, an annulus, to $\Real{}P^2$.
On the other hand, the fundamental group of the adiabatic parameter space is
$\pi_1(\Real{}P^2) = \set{[e], [\gamma]}$, where $e$ is 
homotopic to a point in $\Real{}P^2$ and $\gamma$ a quantity not homotopic to $e$.
We note that $\gamma^2$, i.e., the repetition of $\gamma$ twice is homotopic
to $e$ (see, Fig.~\ref{fig:cover2}~(b)).
Because $\MP$ is a universal cover of $\MB$, there are
only two classes of monodromy actions $\phi_{[C]}$, which is either
$\phi_{[e]}$ or $\phi_{[\gamma]}$. The former and latter cases
correspond to the absence and the presence of the eigenspace anholonomy respectively
(see Fig.~\ref{fig:cover2}~(b) again).
We conclude that $\Phi=\pi_1(\Real{}P^2)$ 
is equivalent with $\Integer_2$.
%TC4 which coincides with the example in \S~3.

% In the present case, $\MP$ covers $\MB$ doubly, and is a universal cover of $\MB$, i.e., $\pi_1(\MP) = \set{[e']}$, where $e'$ is homotopic to a point in $\MP$.
% On the other hand, the fundamental group of the adiabatic parameter space is
% $\pi_1(\Real{}P^2) = \set{[e], [\gamma]}$, where $e$ is 
% homotopic to a point in $\Real{}P^2$, and $\gamma$ is not homotopic to $e$.
% We explain $\gamma$, which is not homotopic to $e$, using an example.
% Imagine a cycle $\gamma$ depicted in the northern unit hemisphere.
% $\gamma$ starts from a point at the equator and arrives at 
% the antipodal point of the initial point. 
% This ensures that $\gamma$ is closed. 
% Furthermore, we suppose that $\gamma$ does not touch 
% the equator on the way. Hence $\gamma$ is not homotopic to $e$ 
% (Fig.~\ref{fig:cover2}~(b)).
% On the other hand, $\gamma^2$, which is the repetition 
% of $\gamma$ twice,
% is homotopic to $e$.

%}%\AT3%
%%%%%%%%%%%%%%%%%%%%%%%%%%%%%%%%%%%%%%%%%%%%%%%%%%%%%%%%%%%%%%%%%%%%%%%%%%%%%

%AT2: add figure and revise the caption accordingly.

\section{Nonadiabatic extension}
%{\AT3%

%Thus far, we have completed our geometrical view of the eigenspace anholonomy 
%with the help of 
In our geometrical treatment of the eigenspace anholonomy so far, there has been no mention
on the scale of temporal variations for the director $\bvec{n}$, nor for the set of projectors $b$. 
In the adiabatic regime, we regard $\bvec{n}$ and $b$ as adiabatic parameters.
%
% In the previous section, $\bvec{n}$ covers the whole non-degenerate
% two-level systems.
%
%We here point out another important role of $\bvec{n}$ and $b$.
%The distinctive point of $\bvec{n}$ and $b$ from other adiabatic parameter is that 
We can also consider the case in which $\bvec{n}$ and $b$ are quantum dynamical variables.
In this case, the time evolution defined by quantum theory naturally
induces the time evolution of $\bvec{n}$ and $b$ which may or may not be adiabatic.
We can then define the cycles made out of nonadiabatic time evolution of $\bvec{n}$ and $b$.

%We here explain an example of nonadiabatic anholonomy. 
We give an example of a two level system described by a time-periodic Hamiltonian, 
whose time dependence may or may not be adiabatic.
%{\TC4 we pose no restriction on.} 
Let $\hat{F}$ denote a Floquet operator, which describes the time evolution during its period. We suppose that $\hat{F}$ has two eigenvectors $\ket{0}$ and $\ket{1}$. The trajectory of the Bloch vector of 
$(\ketbra{0}{0},\ketbra{1}{1})$ 
forms a closed curve, and the trajectory of director of 
$\set{\ketbra{0}{0}, \ketbra{1}{1}}$ is homotopic to a point in $\Real{}P^2$,
implying the absence of anholonomy.

We show that the nonadiabatic anholonomy occurs 
if $\hat{F}$ has two eigenvalues
$\exp\{-i(\epsilon \pm \pi/2)\}$, where $\epsilon$ is an arbitrary real 
number.
%For example, we suppose
Let us write two eigenstates of ${\hat F}$ as $\ket{0}$ and $\ket{1}$, namely
\begin{align}
  \label{eq:eigF}
  \hat{F}\ket{0}
  = e^{-i(\epsilon + \pi/2)}\ket{0},
  \quad
  \hat{F}\ket{1}
  = e^{-i(\epsilon - \pi/2)}\ket{1} .
\end{align} 
A pair of normalized vectors
\begin{align}
  \label{eq:defpm}
  \ket{\pm}\equiv \frac{1}{\sqrt{2}}(\ket{0}\pm\ket{1})
\end{align}
exhibit nonadiabatic anholonomy. This is 
because, from Eq.~\eqref{eq:eigF}, 
$\hat{F}\ket{+} = e^{-i\left(\epsilon+{\pi}/{2}\right)}\ket{-}$
and $\hat{F}\ket{-} = e^{-i\left(\epsilon+{\pi}/{2}\right)}\ket{+}$ hold.
In other words, the trajectory of the Bloch vectors of 
$(\ketbra{+}{+},\ketbra{-}{-})$ draws an open curve during the period
of the Hamiltonian. The corresponding trajectory of the director
must be closed and homotopic to $\gamma$ in  $\Real{}P^2$.

\begin{figure}%[h]
  \centering
  % 3.5cm \simeq 0.23\textwidth 
  \includegraphics[width=3.2cm]{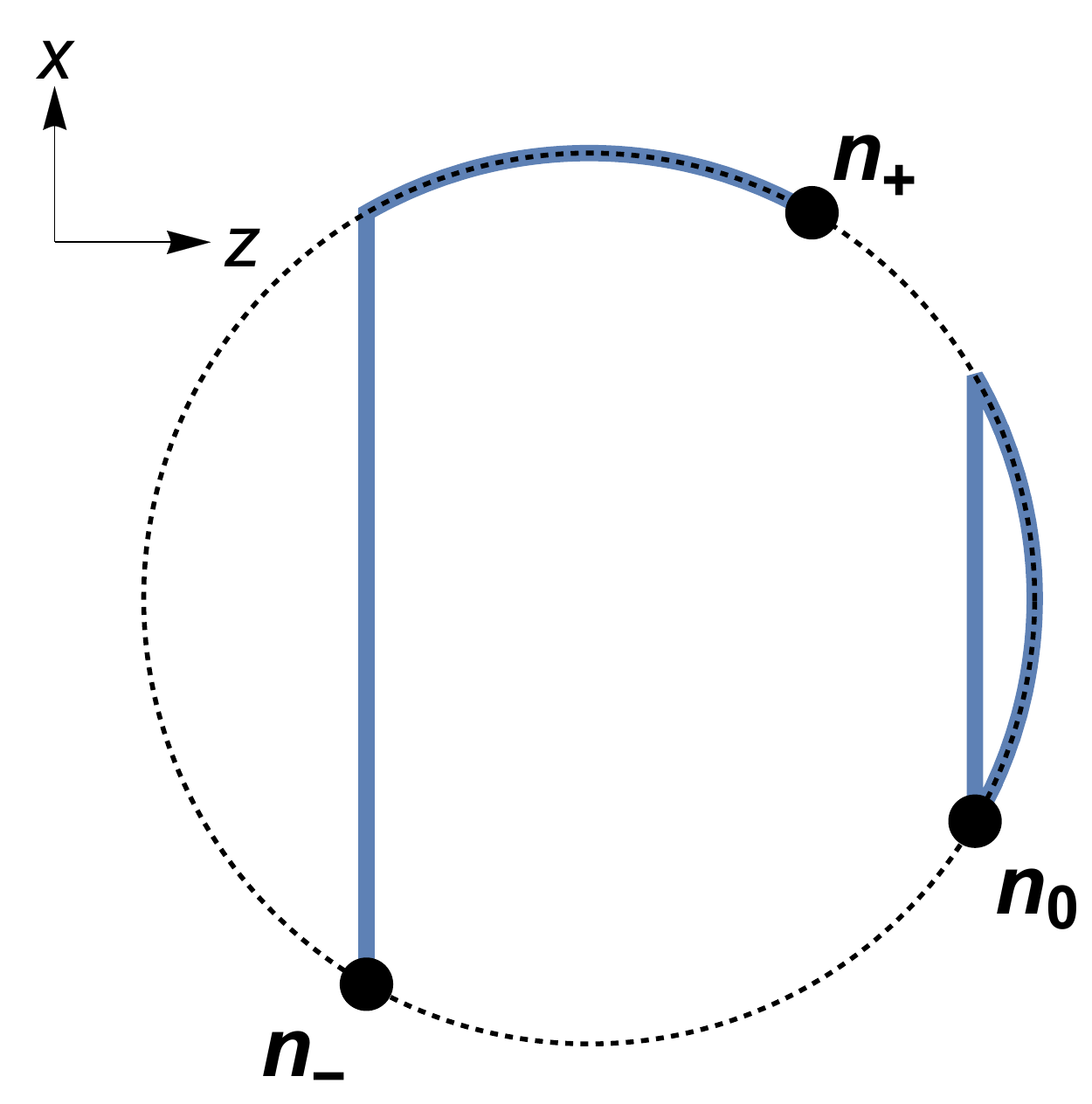}  
  \caption{
    %{\AT3%
      Nonadiabatic cycles in the director space $\Real{}P^2$
      for the kicked spin (Eq.~\eqref{eq:anotherKickedSpin})
      with $B=\pi/3$.
      Trajectories of director is depicted in $(n_z,n_x)$-plane.
      The unit circle in $(n_z,n_x)$-plane is also shown (dotted).
      $\bvec{n}_0$ correspond to 
      the set of eigenprojectors $\set{\ketbra{0}{0}, \ketbra{1}{1}}$
      of $\hat{F}$. The trajectory $C_0$ that starts from $\bvec{n}_0$ 
      is homotopic
      to a contractable loop $e$ (see, Fig.~\ref{fig:cover2}(b)).
      On the other hand, let $C_+$ be the cycle whose initial point
      $\bvec{n}_{+}$ correspond to 
      the set of projectors $\set{\ketbra{+}{+}, \ketbra{-}{-}}$
      (see, Eq.~\eqref{eq:defpm}). The end point of $C_+$ 
      is $\bvec{n}_{-}$, which is equivalent with $\bvec{n}_{+}$. 
      Note that $C_+$ is not homotopic to $C_0$.
      Suppose the system is initially at $\ket{+}$. Due to the time evolution 
      along $C_+$, the system arrives at $\ket{-}$ to exhibit the nonadiabatic
      anholonomy.
    %}%\AT3%
  }
  \label{fig:nonad}
\end{figure}

%{\AT4
An example of $\hat{F}$ is provided by 
%}%\AT4
% {\ATDEL4%
% %{\TC4 
% Another example is
% } 
the kicked spin-$\frac{1}{2}$,
%{\AT4%
which is 
%}%\AT4
described by
the following Hamiltonian
\begin{align}
  \label{eq:anotherKickedSpin}
  \hat{H}(t)\equiv 
  \frac{\pi}{2}\hat{\sigma}_z
  +\frac{B}{2}\hat{\sigma}_y\sum_{n=-\infty}^{\infty}\delta(t-n)
  ,
\end{align}
where we assume $0 < B < \pi$. The period of $H(t)$ is unity.
The corresponding Floquet operator is
$\hat{F}=
\exp(-i \pi\hat{\sigma}_z/2)\exp(-i B\hat{\sigma}_y/2)$,
whose normalized eigenvectors are
% $\ket{0} =\cos(B/4)\ket{\uparrow}-\sin({B}/{4})\ket{\downarrow}$
% and $\ket{1} =\sin(B/4)\ket{\uparrow}+\cos({B}/{4})\ket{\downarrow}$.
\begin{align}
  \label{eq:AKSeigenvectors}
  \ket{0} \equiv
  \begin{pmatrix}
    \cos\frac{B}{4}\\{}-\sin\frac{B}{4}
  \end{pmatrix}
  ,\quad
  \ket{1} \equiv
  \begin{pmatrix}
    \sin\frac{B}{4}\\{}\cos\frac{B}{4}
  \end{pmatrix}
  .
\end{align}
Indeed, they satisfy~Eq.~\eqref{eq:eigF} with $\epsilon=0$.  We depict
the trajectories of the directors of two sets of projectors
$\set{\ketbra{0}{0}, \ketbra{1}{1}}$ and
$(\ketbra{+}{+},\ketbra{-}{-})$ during the unit time interval of the
Hamiltonian~\eqref{eq:anotherKickedSpin} in $\Real{}P^2$
in Figure~\ref{fig:nonad}.
%}%\AT3%

\section{$N$-dimensional extension}
%
%TC
We close our argument by outlining
the extension to cases that involves an arbitrary number, say $N$, of levels.
%TC

%{\AT3%
First, let us note that 
the definitions of $p$ and $b$ (Eqs.~\eqref{eq:p_def} and~\eqref{eq:b_def}),
which consist of mutually orthogonal projection operators, 
and also the definitions of $\MP$ and $\MB$
are applicable to an arbitrary $N$.
We do not attempt, however, to extend the concepts of Bloch vectors 
and directors for now.
%though, for now, there is no extension 
%of the concepts of Bloch vectors 
%and directors for our purpose.
%}%\AT3%
%{\AT3%
%It is worth to 
We emphasize that $\MP$ is a covering space of 
$\MB$~\cite{Lee-ITT-2011,MehriDehnavi-JMP-49-082105}, 
irrespective of the value of $N$.
%}%\AT3%
%We also remark that 
%{\AT3%
The quantity $\MP$
%the $p$-space 
%}%\AT3%
is referred to as a flag manifold~\cite{Adelman-FP-23-211,FlagText}.

%{\AT3%
%On the other hand, for now,
It is still an open problem to provide a complete classification of the cycles for
an arbitrary $N$-level system. This is due to the difficulty in identifying the
fundamental groups of $\MP$ and $\MM$ for the $N$-level case.

Non the less, it is possible to provide a working example of the present 
formulation with an arbitrary $N$ if we choose appropriate sub-family of 
the general $N$-level systems. %AT4
%This correspond to an extension of the analysis shown at the end of \S~3
%(see, Fig~\ref{fig:cover2}(a)). 
Let us examine a family of
quantum maps whose unit time evolution operator is given by
$
\hat{U}(\lambda) \equiv \hat{U}_0\exp(-i\lambda\ketbra{v}{v}),
$
in which $\lambda$ is an adiabatic parameter, 
$\hat{U}_0$ a nondegenerate unitary operator, and 
$\ket{v}$ a normalized vector.
We also assume that $\ket{v}$
is not an eigenvector of $\hat{U}_0$. 
The following is a summary of 
the result of Ref.~\cite{Miyamoto-PRA-76-042115}
on the exotic quantum holonomy in this model.
Let $\ket{n}$ denote the
$n$-th eigenstate of $\hat{U}_0$, where the quantum number $n$ is 
assigned in the increasing order of quasienergy.
%so as to ensure the quasienergies are sorted.
Since $\hat{U}(\lambda)$ is $2\pi$-periodic in $\lambda$, 
the adiabatic parameter space $\MM$
is identified with a circle $S^1$.
Let $C$ denote an adiabatic cycle where $\lambda$ is increased
from $0$ to $2\pi$.
It is shown in Ref.~\cite{Miyamoto-PRA-76-042115} that 
the eigenspace anholonomy is induced by $C$ because
$C$ delivers $\ket{n}$ to $\ket{n+1}$.

%The present formulation is applied to this model.First, 
Applying the present formulation to this model, we first note that
we have $\pi_1(\MM) = \{[e], [\alpha], [\alpha^2], \ldots\}$,
where $e$ and $\alpha$ are a point and a cycle in $\MM$, as
$\MM$ is regarded as $S^1$~\cite{Nakahara-GTP-1990}.
Also we find
$\pi_1(\MP) = \{[e'], [\beta], [\beta^2], \ldots\}$,
where $e'$ is homotopic to a point, and $\beta$ encloses 
the ``hole'' in $\MP$ (cf. Figure~\ref{fig:cover2}(a)).
The $N$ dependence appears in the fact that
the projection of $\beta$ is homotopic to $\alpha^N$.
Hence the isotropy group is 
$H=\{[e], [\alpha^N], [\alpha^{2N}], \ldots\}$.
We conclude
$\Phi(=\pi_1(M)/H)$ is equivalent with $\Integer_N$, 
$N$-th cyclic group.
%}%\AT3%

\section{Discussion}
%TC
We discuss the relationship 
of this work to the previous works
on the eigenspace anholonomy.
In the present formulation, 
%TC
we have identified 
%TC the bottom layer 
the layers in the hierarchy of
the quantum holonomy, which consists of the $b$-space, $p$-space and the
space consists of frames $(\ket{\psi_0},\ket{\psi_1},\dots)$.
%{\AT3%
A
%}%\AT3%
lift of a closed cycle of $b$ to the $p$-space involves 
the anholonomy in $p$. 
%{\AT3
In contrast, a
%A 
%}%\AT3
lift of a path in $p$-space to the frame space
involves the phase holonomy. In particular, Aharonov-Anandan phase is
induced by the cycle in $p$-space. On the other hand, the phase holonomy
associated with a open path in $p$-space correspond to the off-diagonal
geometric phase~\cite{Manini-PRL-85-3067,Mukunda-PRA-65-012102}. 
In a gauge theoretical approach introduced in Ref.~\cite{Cheon-EPL-85-20001},
%TC
the eigenspace anholonomy and the off-diagonal geometric phase are treated together.
%TC
These two concepts are disentangled and assigned to the different layers through 
the present formulation.

%AT2
%TC
A remark is due to the relationship between the exotic quantum holonomy and 
%AT2
Kato's exceptional point (EP), which is a branch point of the Riemann 
surface of eigenenergies, in non-Hermitian quantum 
theory~\cite{biorthogonal,KatoExceptionalPoint}.
%EPs. 
The adiabatic time evolution under the presence of the eigenspace anholonomy resembles a parametric evolution that encloses an EP, in the sense that these evolutions permutate eigenspaces. 
An analytic continuation of adiabatic cycle in Hermitian 
Hamiltonian and unitary Floquet systems
has enabled to interpret the exotic quantum holonomy as the result of 
parametric encirclement of EP in the complex plane~\cite{Kim-PLA-374-1958,Tanaka-JPA-46-315302}.
Although such a correspondence is valid only when an analytic continuation of the adiabatic cycle is available, the topological formulation is applicable regardless of the analytic continuation.
% AT2
% For now, this relationship can be established only when the analytic 
% continuation of adiabatic parameters is available.
Also, we do not know how the 
non-adiabatic extension of exotic quantum holonomy can be associated with EPs.
On the other hand, we remind the readers
that the relationship between 
%AT2
the phase holonomy
%the exotic quantum holonomy
and EPs is established through 
the analysis of the Riemann surface of 
(quasi-)eigenenergy~\cite{MehriDehnavi-JMP-49-082105}.
Because the covering space structure 
naturally
resides in the Riemann surfaces,
an extension of the present approach to non-Hermitian systems 
should be of interest.

\section*{Acknowledgments}
%\ack
AT wish to thank Professor Ali Mostafazadeh for a useful conversation
and Professor Akira Shudo for comments.
This research was supported by the Japan Ministry of Education, Culture, Sports, Science and Technology under the Grant number 24540412.

% iop
%\section*{References}   %NOPREPRINT%

%% for preparation, I prefer to use BibTeX
%\bibliography{local,atsnapshot-diet}

%% for submission, we need to embed the .bbl file here
%\input{local-fixed.bbl} % a workaround

%% End of local-fixed.bbl %%%%%%%%%%%%%%%%%%%%%%%%%%%%%%%%%%%%%%%%%%%

\end{document}